\documentclass[a4paper]{article}

\usepackage[english]{babel}
\usepackage[utf8x]{inputenc}
\usepackage[T1]{fontenc}

\usepackage[a4paper,top=3cm,bottom=2cm,left=3cm,right=3cm,marginparwidth=1.75cm]{geometry}

\usepackage{amsmath}
\usepackage{graphicx}
\usepackage[colorinlistoftodos]{todonotes}
\usepackage[colorlinks=true, allcolors=blue]{hyperref}

\title{Viewing quantum mechanics through the prism of electromagnetism}
\author{Ankit Pandey, Bill Poirier, Luis Grave-de-Peralta}

\begin{document}
\maketitle

\begin{abstract}
In this paper, we demonstrate novel relationships between quantum mechanics and the electromagnetic wave equation. In our approach, an invariant interference-dependent electromagnetic quantity, which we call ``quantum rest mass'', replaces the conventional role of the inertial rest mass. In the ensuing results, photons, during interference, move slower than the speed of light in vacuum, and possess de Broglie wavelength. Further, we use our electromagnetic approach to examine double-slit photon trajectories, and to arrive at the Schrodinger equation's results for a particle in an infinite square well potential.
\end{abstract}

\section{Introduction}

\begin{figure}[h]
\centering
\includegraphics[width=0.3\textwidth]{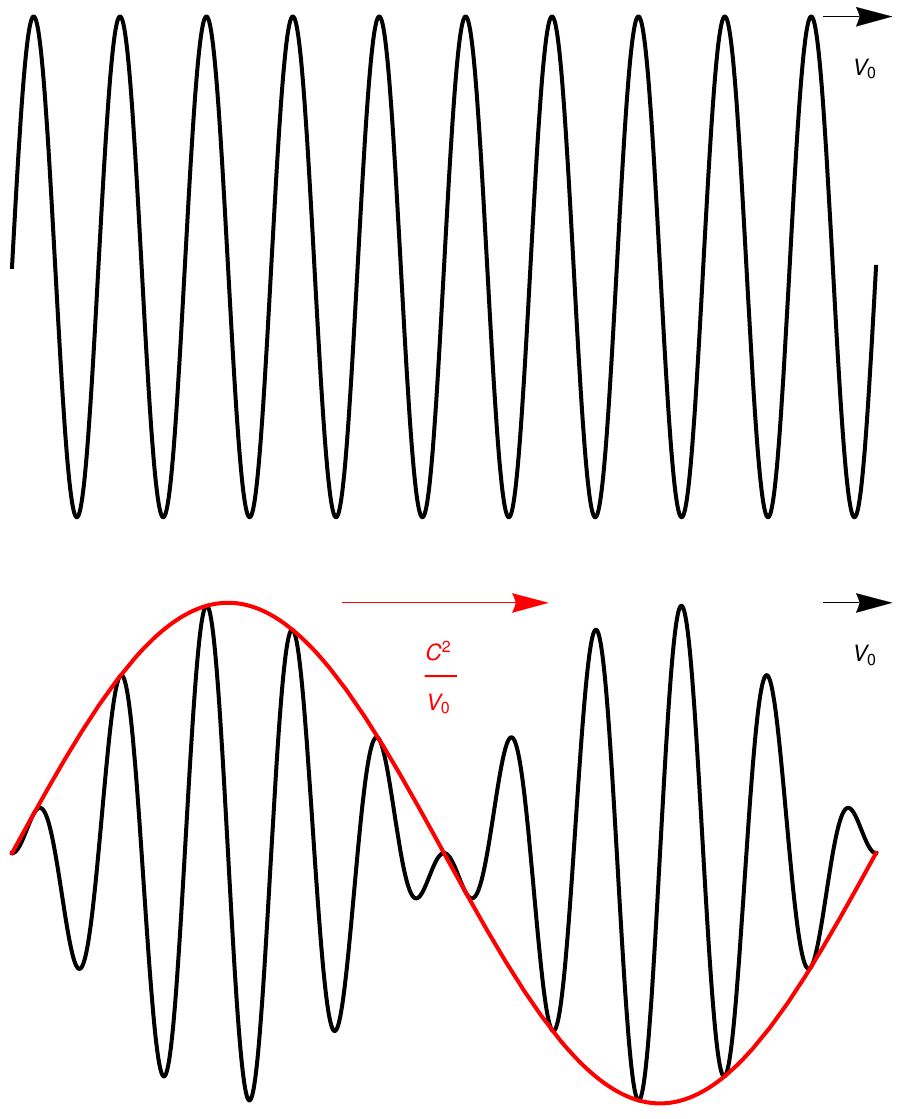}
\caption{\label{fig:frog}The figure at the top represents an electromagnetic standing wave. In the bottom figure, we see the result of a Lorentz-boost acting upon it. Apart from the intuitive forward motion corresponding to the velocity of the boost, the wave also gets modulated by a superluminal wave, shown in red.}
\end{figure}

 We begin this discussion with the following question. What does a standing wave look like when it is boosted in a given direction? One could expect the intuitive answer that we should simply see a standing wave moving in the direction of the boost. This, indeed is the correct answer if we are talking about a Galilean boost of a non relativistic standing wave. But what if the question was about an electromagnetic wave that was Lorentz-boosted? The answer, in this case, is shown in Fig \ref{fig:frog}. \newline

Interestingly, the standing wave gets modulated by a long superluminal (faster than light) wave when seen from the boosted frame of reference. This superluminal wave is more widely known in the context of phase waves obtained from the Klein-Gordon equation \cite{KG1}. If we assume that the standing wave was created by a single photon of frequency ($\omega$) oscillating inside a cavity, then the superluminal wavelength is found to obey:
\begin{equation}
\lambda_{sup}=\frac{h}{m' v}
\label{eq:supLength}
\end{equation}
where 
\begin{equation}
\label{eq:mp}
m'=\frac{E}{c^2}=\frac{\hbar \omega}{c^2}
\end{equation}\newline

Now this formula looks like the de Broglie wavelength equation, but the quantity $m'$ here is not related to the inertial rest mass; the latter is always zero for light. However, rest mass can be defined in more ways than one. We are going to start out by defining it in a way which yields the value shown in Eq. \ref{eq:mp} for standing waves of light.  

\section{Defining ``quantum rest mass'' for light}
In general, rest mass can be derived from the momentum four vector as follows \cite{relativity}:
\begin{equation}
\label{eq1}
   m^2 = \frac{P_{\rm \mu}P^{\rm \mu}}{c^2}
\end{equation}We note that if standing waves of light are assumed to correspond to a particle being at rest, then the particle mass indeed corresponds to Eq. \ref{eq:mp}:
\begin{equation}
\label{lightMomentum}
\begin{split}
        &P^{\rm \mu}= \left(\frac{\hbar \omega}{c}, 0, 0, 0 \right)
\end{split}
\begin{split}
        & \Rightarrow m^2= \frac{\hbar^2 \omega^2}{c^4}
\end{split}
\end{equation}
\begin{equation}
\Rightarrow m=\frac{\hbar \omega}{c^2}
\label{eq:mass}
\end{equation} \newline

  Standing waves of light are the superposition of a forward-moving and a backward-moving sinusoidal wave. The average momentum of these two sinusoidal waves is zero. By defining standing wave photons to be at rest, we are associating photon trajectories with this average momentum. To distinguish $m$ from inertial rest mass, we shall call it the ``quantum rest mass''. \newline

  More generally, we can consider two waves of equal amplitude but different frequencies $\omega_+$ and $\omega_-$, moving in opposite directions. This superposition is what we call a ``bidirectional wave''. It should be understandable that there will exist a boosted frame of reference $f'$ in which these frequencies are equal. To calculate the quantum rest mass of photons in this case, one can move to the $f'$ frame of reference and use Eq. \ref{lightMomentum}. Alternatively, it can be shown that one can also calculate the quantum rest mass of a bidirectional wave from an arbitrary frame of reference by simply using Eq. \ref{gss}, where the wave is assumed to live in the x-axis.
\begin{equation}
 \label{gss}
    P^{\rm \mu}= \frac{\hbar}{2c}\left( \omega_++\omega_- , \omega_+-\omega_-, 0, 0 \right)
\end{equation}\newline

  It should be noted that the quantum rest mass of photons depends on interference. Where there is no interference, the quantum rest mass of photons is zero, just like the inertial rest mass.  To drive home this point, we explain the double slit pattern in the 2-dimensional space with our approach. 

\section{Double slit interference}
In the double slit experiment, two wavefronts intersect each other. The angle of intersection varies at each point $(x,y)$ of space. Let us say that their wave vectors intersect at the angle $\theta(x,y)$. The following equations describe the quantum rest mass of the photon, and its corresponding velocity, ``locally'' at $(x,y)$:
\begin{equation}
m(\theta)=\frac{\hbar\omega}{c^2}\sin{\left(\frac{\theta}{2}\right)}
\label{eq:localMass}
\end{equation}
\begin{equation}
v(\theta)=c \cos{\left(\frac{\theta}{2}\right)}
\label{eqLocalV}
\end{equation}
The local velocity points in the direction mid-way between the individual wave vectors. Note that theta can only vary between 0 and $\pi$.  The ``local wavelength'' corresponding to the above quantities, given by
\begin{equation}
\lambda_{sub}(\theta)=\frac{h}{m(\theta) v(\theta)},
\label{eqLocalLambda}
\end{equation}
lies in the direction perpendicular to the local velocity.  \newline

\begin{figure}[h]
\centering
\includegraphics[width=0.5\textwidth]{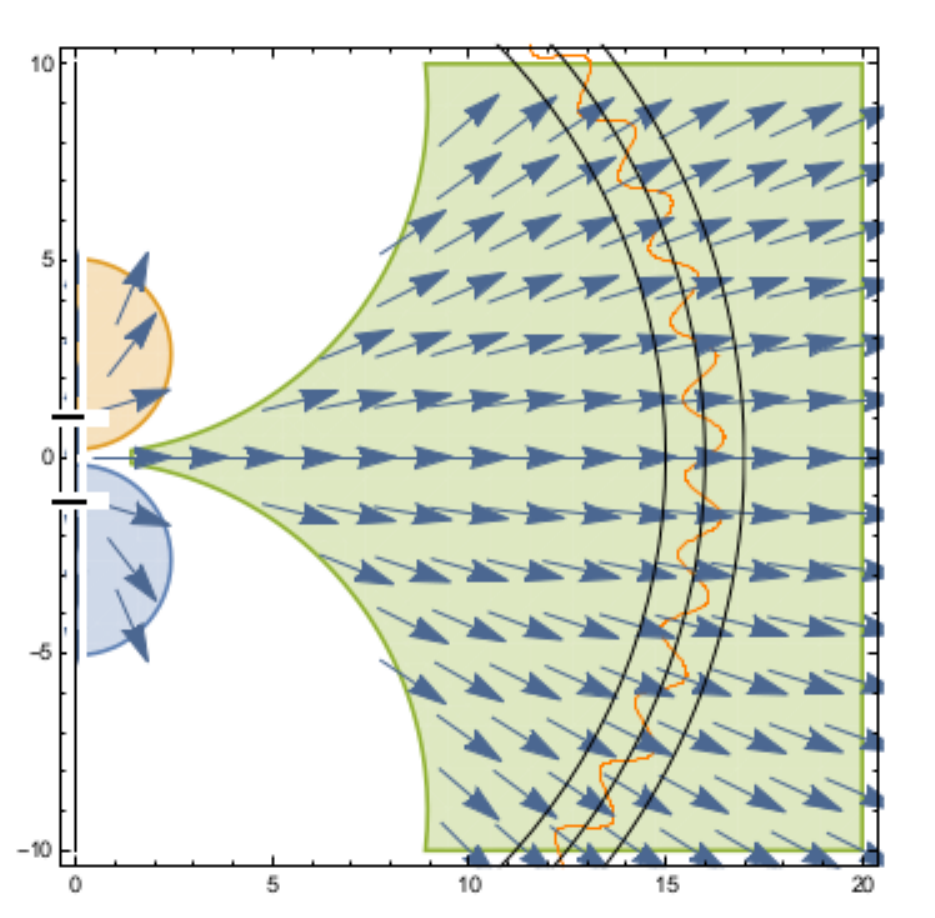}
\caption{\label{fig:DS}Double slit interference can be studied using quantum rest mass of light, as explained in the text.}
\end{figure}

The amplitude contributed by a given slit decays by $\frac{1}{r}$ as a function of distance $r$ from a given slit. Due to this, it is understandable that in regions very close to either of the two slits (the regions represented by red and blue colors in Fig \ref{fig:DS}), the contribution of the other slit is comparatively negligible. The trajectories here move radially outwards from the two slits, and have zero quantum rest mass. On the other hand, in the green region, the amplitudes contributed by both the slits are rather similar. Here, Eq. \ref{eqLocalLambda} holds true, and thus, the trajectories point radially outward from the midpoint between the slits, as can be derived from the comment below Eq. \ref{eqLocalV}. \newline

Please keep in mind that $\theta$ is not the angle extended at the origin of Fig \ref{fig:DS}, but the angle between the wave vectors intersecting at a given point. It remains relatively constant for small movements perpendicular to the velocity vector, for example, along the arc shown in the figure. \newline

Let us use the quantum rest mass to find the spacing between two bright fringes on a screen far away from the slits.  For small vertical displacements along this arc from its center in the figure, we can approximate:
\begin{equation}
\sin{\frac{\theta}{2}}=\frac{d}{2D}
\end{equation}
where D is the distance of the arc from the origin, and d is the distance between the slits. We can also approximate $\cos{\frac{\theta}{2}}=1$. The distance between two consecutive bright fringes along the vertical axis is half the local wavelength, which is calculated from Eq. \ref{eqLocalLambda}. It is found to be equal to
\begin{equation}
\frac{\lambda_{sub}(\theta(0,D))}{2}=\frac{D\lambda}{d}
\end{equation}Where $\lambda=\frac{2 \pi c}{\omega}$. This result is the same as that obtained by the traditional interference method. \cite{optics}
\newline

The above method can be used further to numerically draw massive photon trajectories or to plot the photon quantum rest mass as a function of space. It is easily noted that the trajectories are the most massive (in terms of quantum rest mass) around the region where the red, blue and green regions in Fig \ref{fig:DS} are closest to each other.

\section{Particle in a box solution via bidirectional waves}
\begin{figure}[h]
\centering
\includegraphics[width=0.5\textwidth]{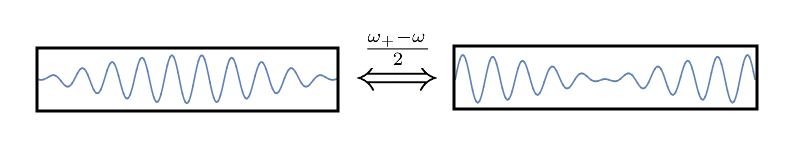}
\caption{\label{imageOscillation} The ``lower frequency'' mentioned in the text corresponds to slow oscillations of the waveform between a broad sine function and a broad cosine function. The frequency of these oscillations is mentioned in the figure.}
\end{figure}

In this section, we derive a Schrodinger-like \cite{qm} ``particle in an infinite square well'' solution in 1 dimension using bidirectional waves. Physically, we aim to model a massless cavity, which in turn is placed inside a much larger infinite square well potential. The cavity, representing a particle, is assumed to move with the velocity v in either direction. To model this situation with electromagnetic waves, we consider the superposition of two bidirectional waves moving with the velocity v opposite to each other inside an infinite square well. Let us call this superposition F(x,t). When this function is plotted and animated with time, two distinct frequencies can be visually observed: A higher frequency causing standing-wave-like oscillations, and a much lower frequency. The latter frequency causes slow broad oscillations of the waveform between being sine-like and cosine-like, as shown in Fig \ref{imageOscillation}
\newline
It should be kept in mind that the ``particle'' is not represented by the waveform itself, but by the cavity. Although we have drawn the waveform of light all along the infinite square well, we need to remember that the photon is constrained within the cavity. Let us assume that the cavity is of length L. If the cavity's instantaneous location is assumed to be at an arbitrary location x, then the waveform only exists within the boundary of the cavity, i.e, between $x-L/2$ and $x+L/2$. Nevertheless, the waveform \emph{inside this region} will be the same as F(x,t) in this region. Hence, F(x,t) should be thought of as the ``internal structure of the particle", as opposed to being an analogy of its Schrodinger-like wavefunction. \newline

\begin{figure}[h]
\centering
\includegraphics[width=0.5\textwidth]{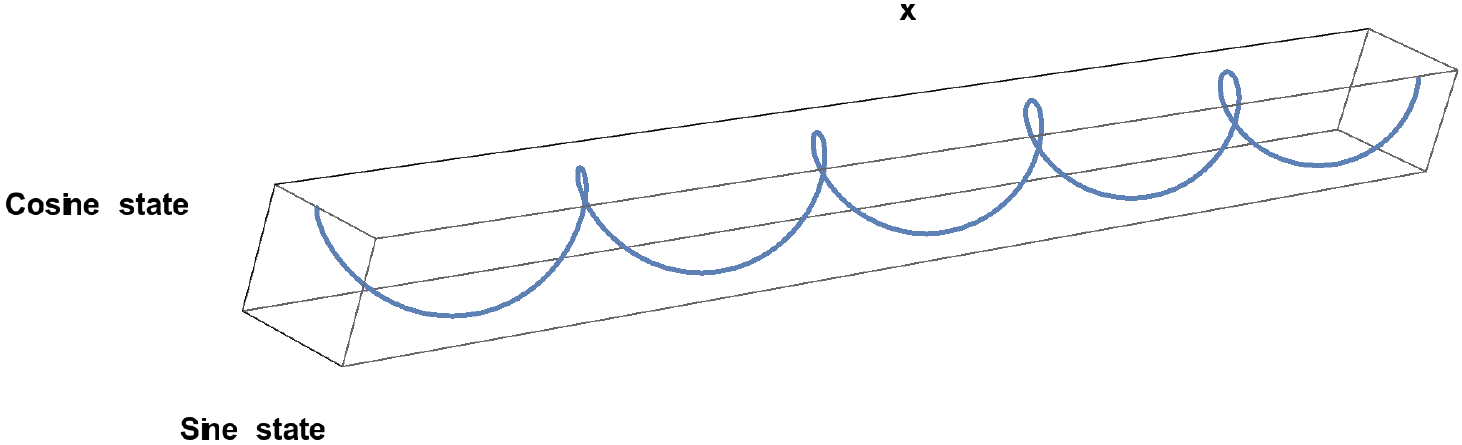}
\caption{\label{imageSinCos}The horizontal axis represents the displacement (x) of the cavity. The two axes perpendicular to the x-axis represent the amplitudes of the Sine and Cosine states as a function of x. }
\end{figure}

To derive a Schrodinger like wavefunction from F(x,t), we can proceed as follows. Consider the two functions between which the system oscillates in Fig \ref{imageOscillation}. We shall hereby call these functions as its two ``internal states'', namely the Sine state and the Cosine state.  For simplicity, we also assume that the cavity is much smaller than the well.  In Fig \ref{imageSinCos}, we plot the amplitude of these states as a function of the cavity's position as it moves with its constant velocity v. It is found in this figure that the wave has the correct de Broglie wavelength that is expected from the quantum mass $m$. Note that the sine and cosine states here bear resemblance to the real and imaginary components of the Schrodinger's wavefunction of a particle. Thus, it can be easily observed that by superimposing  Fig \ref{imageSinCos} with the corresponding function for the cavity moving in the opposite direction, and requiring that the superimposition vanishes at the boundaries, we recover the Schrodinger equation's  solutions for particle in an infinite square-well.

\section{Concluding remarks}
The quantum rest mass of light, as defined in this paper, provides alternative insights into various electromagnetic phenomena involving the interference of light.  Beyond electrodynamics itself, this concept also helps provide unique approaches towards understanding and appreciating quantum mechanics from a relativistic semi-classical perspective. \newline

\section{Acknowledgements}
B. Poirier acknowledges a grant from the Robert A. Welch Foundation (D-1523). We also thank Igor Volobouev and Maik Reddiger for their valuable opinions and discussions on these topics.

\end{document}